\documentclass[acmtog]{acmart}

\def\BibTeX{{\rm B\kern-.05em{\sc i\kern-.025em b}\kern-.08emT\kern-.1667em\lower.7ex\hbox{E}\kern-.125emX}}
    
\usepackage[autostyle=true,german=quotes]{csquotes}
\usepackage{makecell}
\usepackage{array} 
\usepackage{amsmath,graphicx}
\renewcommand\footnotetextcopyrightpermission[1]{}
\setcopyright{none}
\makeatletter
\renewcommand\@formatdoi[1]{\ignorespaces}
\makeatother
\begin{document}

%
\title{Co-sleep: Designing a workplace-based wellness program for sleep deprivation}

%

\author{Bing Zhai}
\affiliation{%
  \institution{Open Lab, Newcastle University}
  \city{Newcastle upon Tyne}
}

\author{Stuart Nicholson}
\affiliation{%
  \institution{Open Lab, Newcastle University}
  \city{Newcastle upon Tyne}
}

\author{Kyle Montague}
\affiliation{%
  \institution{Open Lab, Newcastle University}
  \city{Newcastle upon Tyne}
}

\author{Yu Guan}
\affiliation{%
  \institution{Open Lab, Newcastle University}
  \city{Newcastle upon Tyne}
}

\author{Patrick Olivier}
\affiliation{%
  \institution{Monash University}
  \city{Melbourne}
}

\author{Jason Ellis}
\affiliation{%
  \institution{Northumbria University}
  \city{Newcastle upon Tyne}
}
%
\begin{abstract}
Sleep deprivation is a public health issue. Awareness of sleep deprivation has not been widely investigated in workplace-based wellness programmes. This study adopted a three-stage design process with nine participants from a local manufacturing company to help raise awareness of sleep deprivation. The common causes of sleep deprivation were identified through the deployment of technology probes and participant interviews. The study contributes smart Internet of things(IoT) workplace-based design concepts for activity tracking that may aid sleep and explore ways of sharing personal sleep data within the workplace. Through the use of co-design methods, the study also highlights prominent privacy concerns relating to use of personal data from different stakeholders' perspectives, including the unexpected use of sleep data by organisations for fatigue risk management and the evaluation of employee performance. The Actigrahy and sleep diary data can be accessed online through {\url{https://github.com/famousgrouse/pervasivehealth/}} 
\end{abstract}

%
%


%

%

%
\maketitle
\thispagestyle{empty}
\section{Introduction}
Sleep deprivation is a growing public health issue. Within the UK, approximately one-third of the population suffer from sleep deprivation to varying degrees, according to research published by the National Health Service (NHS) in 2011\cite{bazian2011sleep}. Half of the population averages only six hours or less sleep, per night, which contrasts dramatically with a figure of only 8\% in 1942\cite{cooke2017sleep}. Lack of sleep significantly impacts on people's work and overall quality of life. People typically spend 30\% of their time at work with another 30\% or less time asleep. This sleep-loss epidemic has been estimated to cost the UK \pounds40 billion annually, which equates to around 200,000 lost working days each year due to insufficient sleep; this is the equivalent to 1.9\% of the GDP \cite{Hafner2016lack}.

Sleep deprivation over a sustained amount of time negatively impacts on people's lives causing daytime drowsiness and impaired cognitive function\cite{Durmer2005neurocong}. Moreover, it can increase the risk of more severe health conditions, such as heart disease, high blood pressure and diabetes\cite{Cappuccio2010sleep}. A recent sleep study demonstrated that less than 4 hours of sleep per night is similar to the effects of excess alcohol consumption \cite{Roehrs1994sleepiness,Williamson2000moderate}. Furthermore, chronic sleep deprivation studies suggest that less than 7 hours of sleep per night in healthy adults results in diminished cognitive ability, equivalent to the symptoms of individuals who remain awake for 28-48 hours \cite{Van2003thecumulative}. For office workers, this would significantly reduce their efficiency, potentially leading to an increase in work-related stress due to the demands of having to 'catch-up' on daily workloads \cite{Dahlgren2005different}. In general, the risks of sleep deprivation have not been widely disseminated in the UK and public awareness remains low \cite{Altevogt2006sleepdisorder,Cappuccio2010sleep,Konnikova2014agene}.

Information technology can be used to aid sleep research; within HCI, the design of systems that support data collection, self-reflection and self-behaviour change have been addressed by personal informatics \cite{Li2010astage}. Personal informatics tools and mobile applications which support sleep tracking are becoming ubiquitous \cite{Sleepas2018,Choe2015sleeptight}.  For example, a number of mobile apps have been designed to assist people to improve their sleep hygiene behaviours based on a stage-based model \cite{Fritz2014persuasive}. However, when designing for tracking and sharing personal health data, such as in workplace, these introspective models face a number of challenges, including pervasive data capture, and issues pertaining to personal privacy and health data sharing. Whilst researchers have begun to explore the design of personal informatics systems for large-scale work environments that encourage positive behaviour change through workplace-based wellness programs \cite{Chung2017finding,Mattke2013workplace} such as short-term step-counting campaigns \cite{buis2009evaluating,Gorm2016steps} programs or campaigns that encourage employees to sleep well are rare. 

Our three-phase design process lasted for one month and was conducted with a local manufacturing company. Nine participants were provided with activity bands and a sleep diary to track their sleeping patterns over a week. They were invited to an interview where they could analyse and annotate their own sleep data and explore the circumstances surrounding their sleeping patterns as well as investigate their further willingness to track personal activities. We also explored their willingness to collect personal data in design workshops. Discussions then took place to consider how the Internet of Things (IoT) devices could be used within the shared work space for the purposes of automating personal data collection, and to determine the range and granularity of personal data to be shared. 

Finally, a set of interactive dashboards were created to represent the worker's personal data; these were built and used as digital probes. Nine interviews were then conducted which explored people's considerations around data privacy when using the dashboards. Our interviews revealed that sleep data could be used to shape the company's health and safety policies and influence performance management. Further discussions highlighted potential solutions to overcome concerns surrounding people's personal data and its use within an organisation's wellness program. Our findings contribute design considerations for the deployment of a sleep wellness program in respect of workers' concerns surrounding data privacy and organisations potential miss-use of workers' personal data. 

\section{Related work}
\subsection{Sleep deprivation issues in manufacturing sectors}
Sleep deprivation affects the manufacturing industry in the form of diminished productivity, increased product defect rates and increased accident rates. Workers on shifts as part of a 24-hour cycle within manufacturing companies are highly prone to sleep deprivation; this can lead to circadian misalignment which is the primary factor that occurs when the body clock is altered to the extent that someone finds it difficult to fall asleep during night time \cite{magnavita2017sleep}. Insufficient sleep can impair attention to details, alertness, concentration and reasoning. This can also make office worker more likely to make mistakes, especially when they are engaging in repetitive administrative work; this could be further compounded by stress and long-term sedentary nature of sitting at a desk. \cite{Lerman2012fatigue}.

Many companies implement fatigue management plans to prevent accidents. This includes shift managers observing and training workers to maximise daytime sleep opportunities; it also enables employees to stick to a fixed shift pattern rather than a rotation shift. Alternative methods include adopting micro tasks to assess fatigue and prevent accidents as well as adopting flexible working schedules \cite{Lerman2012fatigue}.

\subsection{Digital Innovation in Sleep deprivation}
Choe et al.\cite{Choe2014understanding}  revealed that people have a strong willingness to track their sleep and factors connected to their sleep in a low-burden manner. The study also emphasised that long-term tracking of sleep is also vital for people to identify the trends to assist them to build their desired objectives in sleep pattern. Many studies and commercial mobile apps have explored ways of capturing sleep patterns and providing feedback and/or suggestions for sleep hygiene-related behaviours (e.g. avoid use electronics before going to bed) that relate to personal sleep health \cite{Bauer2012shuteye,Choe2015sleeptight,Daskalova2016sleepcoacher,Durmer2005neurocong}. Furthermore, Pina et. al \cite{Pina2017frompersonal} extended a personal informatics system in designing for a family to track and share sleep data among its members - a first step towards collectively addressing sleep issues that frames interventions as social processes whereby data is shared with family members to understand the causes of lack of sleep. 

\subsection{Design for Sleep Personal informatics \& Smart Workplace}
Personal informatics is a class of tool that helps people collect data for self-monitoring in personal and public space. Recent HCI research has begun to explore personal tracking experiences and its meaning in specific social contexts \cite{Garbett2018think}. For instance, people share their personal data through social media integration or in-app friend networks to compare and compete with friends and colleagues\cite{Epstein2015from}. Moreover, researchers use lived informatics sociability features to promote shared reflection in a group of trackers based on visualised and aggregated personal data; such instances seek to go beyond simple behaviour change\cite{Baumer2015reflectiveinformatics,Garbett2018think,Puussaar2017enhancing,Rooksby-2014-personaltracking}. These studies extend beyond behaviour modification and raise questions about how we should integrate our data into a shared, reflective, and interactive everyday lived experience with privacy protection \cite{Lin2006finishnstep}. 

Personal informatics data can be collected using IoT technologies, that is, the network of physical devices embedded with sensors, software, actuators and electronics that when connected to the internet allows existing computer-based systems to sense the physical world \cite{Fleisch2010what}. Deploying IoT devices in the workplace has the potential to address public health issues - for example, encouraging handwashing in hospital\cite{Staats2016motivating} - through pervasive data collection, and actions were taken in response to the analysis of such data. 

\subsection{Designing for Health Tracking Workplace Wellness Program}
Employee health promotion is an activity that employers are increasingly engaging in [18]\cite{Chung2017finding}. For example, to encourage employees to exercise on their own, many companies subsidize their employees' purchases of fitness trackers and/or gym membership \cite{Gazmararian2013arandom,Mattke2013workplace,Wells2017incorporating}. Activity trackers are personal informatics device most commonly used to collectively encourage behaviour change within the work environment as part of employer health and wellbeing campaigns [18; 32; 33] \cite{Chung2017finding,Gorm2016sharing,Gorm2016steps}. Such programs can be divided into two categories, short-term project and long-term projects.  Short-term projects usually last for one to three months in various forms (individual or team based) and include incentives such as cash reward or gift card to encourage employee participation in events \cite{Gorm2016sharing,Green2007peer}. Within long-term programs, employees also receive the incentives (e.g. virtual points) based on their own health tracking data to maintain participation in the project. These virtual points can be redeemed at the end of the year as cash rewards, gift cards, or insurance discounts. 

\subsection{The Criticisms of workplace surveillance }
The use of tracking technology in workplace wellness programs has not gone unnoticed by the media who have generally adopted a critical view of such practices. With the boundary between life and work being slowly blurred by the use of personal informatics technologies, the question of whether employees have a choice to participate in the use of personal tracking, or are in practice coerced into participating, has attracted significant attention from both journalists and researchers \cite{Datoo2014these,Lupton2014health,Manokha}.  

The increasing quantification of employee activities warrants concern and a factor that workplace wellness programs rarely consider or design from an employee's perspective in this respect \cite{Robinson2015thequantified}. Incentive-based health tracking programs may have implicit risks for employees (who cannot afford to say no), and the very program that is intended to promote their wellbeing has the potential to be a vehicle for their dis-empowerment \cite{Sharon2017selftracking}.The latest developments in sensor technology have reduced the size of devices so that they are even smaller. Devices can now monitor people's biological signals such as breathing, attention and stress levels within different environments, such as the workplace.  environments\cite{oconnor2015wearables}. The work of Sharon \cite{Sharon2017selftracking} expresses real concern about monitoring people's workplace productivity. ~\enquote{While employers usually explain that such programs aim to optimise the workplace by making people healthier and adding a little "friendly competition" to improve morale and productivity, for critics, their disciplining effects are unmistakable; the line between voluntary participation and compulsory participation is not always clear-cut}.

\section{PROJECT DESIGN}
There is a lack of research focus on how to empower employees to understand the relationship between sleep deprivation and their daytime life from a personal and collective perspective through a workplace wellness program. Our research aims to empower employees and to encourage them to participate directly in the construction of their wellness program, as well as let them decide which data should be collected and shared. We aimed to explore a bottom-up approach by co-designing the wellness program with employees to address sleep deprivation specifically. In particular, our goal was to derive design implications for the system that can be used in the workplace to promote reflection and dialogue about sleep deprivation and encourage employees to improve their sleep collectively. Focusing on employees from a local manufacturing company, the project ran over one month with three stages: 1) design and deployment of digital and cultural probes for gathering accounts and understanding sleep deprivation issues; 2) organisation of design workshops for acquiring sleep-related information using IoT technologies in the workplace 3) designing sleep data visualisations and conduct data-driven interviews that incorporated notions of corporate business social responsibility and workplace and personal ethics. 

The first phase of the study aimed to gain an in-depth understanding of the factors related to people's sleep through nine interviews with office workers. The second stage used that understanding to undertake an IoT card-based design workshop. The first purpose of the workshop was to involve users in the design of the smart work space by which data could be automatically using IoT technology. The second purpose of the workshop was to let users decide which data should be shared, and in what ways to visualise the data. The third phase of the study involved the visualisation of data collected in the previous phase on an interactive dashboard which we presented to users in a process of exploring how they understand and would like to use the data.

\subsection{Participant recruitment}
A manufacturing company in England comprising of approximately 25 employees was chosen as being an appropriate case study partner. The main protocol of the study was sent to the company's HR manager and director. In their previous year's strategic plan, employee's welfare and wellness were regarded as a major investment and their business social responsibility. Health and safety in the manufacturing environment have been seen as a priority to be addressed. Nine participants were recruited for this study.

\begin{table}
  \caption{Participant job roles}
  \label{tab:partjobrole}
  \begin{tabular}{|p{3.5cm}|p{3.5cm}|}
    \hline
    \textbf{Participants}&\textbf{Role}\\
    \hline
    P1,P2,P3,P7&Office Worker\\
    P4,P8,P9 & Shift Worker\\
    P5,P6 & Management\\
  \hline
\end{tabular}
\end{table}

\section{Stage one: Understanding the causes of sleep deprivation}
A sleep data profile was gathered using a cultural probe; this consisted of a thirteen page sleep diary and separate log sheets as well as empty data visualisation charts. The paper-based sleep diary was constructed of set questions and allowed people to record sleep-related factors at any time of the day, e.g. if they woke up at any point during the night they could record the reason. The log sheets allowed people to track any other factors that people felt were important to on their sleep. e.g. a bird chirping whilst sitting on the bedroom windowsill. The log sheets also served as an alternative for participants who did not want to wear a wristband, but still wanted to participate in the study. Some participants believed that using a paper-based sleep diary would reduce the negative effects associated with using electronics before bed. 

\subsection{Wearable sensors and sleep diary}
Axivity AX3 sensor wristbands were utilised as digital probes.They are tri-axial accelerometer (Actigraphy) devices designed by Axivity\texttrademark which were used to collect unconscious limb movements during sleep\cite{Doherty2017largescale}. The Ax3 actigraphy equipment has been widely used in ambulatory sleep monitoring studies by clinics\cite{Ancoli2003Therole}. The sleep diary used in this study was adapted from National Sleep Foundation.\footnote{\url{https://www.sleepfoundation.org/articles/nsf-official-sleep-diary}}. It is a commonly used sleep diary. In addition to being able to record sleep onset and offset, it also records questions that related to sleep, more details see Table ~\ref{tab:sleep_diary}

\begin{table}[htb]
  \centering
  \caption{National Sleep Foundation Sleep Diary}
  \resizebox{\columnwidth}{!}{%
    \begin{tabular}{|l|l|}
    \hline
    \textbf{Questions } & \textbf{Answering Format} \\
    \hline
    I went to bed last night at  & \makecell[c]{AM/PM}\\
    \hline
    I got out of bed this morning at  & \makecell[c]{AM/PM} \\
    \hline
    Last night I fell asleep & \makecell[c]{Easily,\\ After some time,\\ With difficulty} \\
    \hline
    I woke up during the night & \makecell[c]{number of times} \\
    \hline
    I woke up during the night & \makecell[c]{number of minutes} \\
    \hline
    Last night I slept a total of  & \makecell[c]{number of hours} \\
    \hline
    My sleep was disturbed by & \makecell[c]{Noise, lights, \\pets, allergies,\\ temperature, \\disconfort, \\stress, etc.}\\
    \hline
    When I woke up for the day, I felt & \makecell[c]{Refreshed,\\ Somwhat refreshed,\\ Fatigued}\\
    \hline
    I consumed caffeinated drinks in the  & \makecell[c]{Morning, \\Afternoon, \\Evening} \\
    \hline
    Medications I took today & \makecell[c]{Any medicines \\ influence sleep }\\
    \hline
    Took a nap? & \makecell[c]{Yes, No} \\
    \hline
    \makecell[l]{During the day, how likely was \\I to doze off while performing \\daily activities  }& \makecell[c]{No chance,\\ Slight chance, \\Moderate chance, \\High chance} \\
    \hline
    Throughout the day, my mood was &\makecell[c]{Very pleasant,\\ Pleasant, \\Unpleasant, \\Very unpleasant} \\
    \hline
    \makecell[l]{Approximatly 2-3 hours before \\going to bed, I consumed}& \makecell[c]{Alcohol, \\A heavy meal,\\ Caffeine,\\ Not applicable} \\
    \hline
    \makecell[l]{In the hour before going to sleep,\\ my bedtime routine included} & \makecell[c]{Reading a book,\\ using electronics,\\ taking a bath, \\doing relaxation \\exercises,etc.}\\
    \bottomrule
    
    \end{tabular}%
    }
  \label{tab:sleep_diary}%
\end{table}%

\begin{table}[hbt!]
  \centering
  \caption{Number of Recorded Valid Sleep Episodes by Actigraphy and Sleep Diary}
   \begin{tabular}{|p{1.9cm}|p{1.6cm}|p{1.6cm}|p{1.5cm}|}
    \hline
      & \multicolumn{1}{l|}{\makecell[c]{Dominant\\ Wrist}} & \multicolumn{1}{l|}{\makecell[c]{Non-dominant \\ Wrist}} & \multicolumn{1}{l|}{\makecell[c]{Sleep \\ Diary}} \\
    \hline
    Valid Episodes & \makecell[c]{50} & \makecell[c]{43} & \makecell[c]{46} \\
    \hline
    \makecell[l]{Mean Valid \\Sleep Episodes}& \makecell[c]{6.25} & \makecell[c]{5.38} & \makecell[c]{5.75} \\
    \hline
    \end{tabular}%

  \label{tab:sleeprecruitement}%
\end{table}%
\subsection{Sleep data collection}
A total of nine people without sleep disorders were recruited. Eight participants wore Axivity 3 wristbands on both dominant and non-dominant wrists while filling out their sleep diary for a week. One participant used only a sleep diary to track sleep as the subject was uncomfortable to wear the wristband. Table ~\ref{tab:sleeprecruitement} shows the details of the valid nights by both methods. We also made our data set available online that included the sleep actigrahy and diary (anonymised).\footnote{\url{https://github.com/famousgrouse/pervasivehealth}} The sleep data was processed by an "Estimation of Stationary Sleep-segments"\cite{borazio2014towards} algorithm to generate actigraphy data which is typically used for determining sleep pattern and circadian rhythms. Afterwards, one-to-one follow-up interviews were conducted with each participant. Each person was given a \pounds25 amazon voucher to compensate their time spend engaging with the study.

The purpose of the interview was to gain insight into the subject's daily routine based on the recorded sleep data to find out: 1) the factors which influenced their sleeping routine; 2) their attitude, opinions and concerns toward the tracking of their personal activities; 3) the configuration of the wellness program (e.g. common objectives, how personal data could be shared within the office etc). A thematic approach was taken in the data analysis\cite{Braun2006usingthematic}; the data was coded inductively and summarised with shortcodes, which were then grouped into larger candidate themes.

\subsection{Interview Findings}
\subsubsection{Lack of awareness about effective sleep}
Of the nine participants, six did not realise that they were lacking sleep before engaging in this study. They thought they had slept for long enough because they counted time drifting on the bed as time being asleep. For example, one participant stated:\enquote{I think I've probably been awake at times when I don't realise I've been awake.}(P2). By cross-comparing the actigraphy data with participants sleep diaries, many participants realised that their sleep time was actually less than seven hours per day. 

Five out of nine participants understood the number of hours they needed and how the amount of sleep can impact on their work. Such as P8 stated:\enquote{If everybody were going into work and had a good eight hours' sleep, I would say that production figures would rise, without a doubt.}   

\subsubsection{Family reasons \& personal reasons}Many of the participants noted that family obligations often interfere with a regular sleeping routine. Those who have children, elderly or who are in a position of care over other family members have more responsibility outside of the work environment. P9 noted: \enquote{My mum's an alcoholic and she's got Alzheimer's. My dad's 71 and just suffered cancer}. Additionally, stress and fatigue can also affect the quality of sleep. Those with children and partners explained that they were sometimes inadvertently woken up when sleeping as P8 explains: \enquote{Whether you like it or not, she wakes you up because she's noisy when she (wife) comes in.}

Other factors that affect sleep included water and caffeine intake (P6), the amount of noise, light (P4, P7) and the temperature (P2). The activities of household pets and animals can also cause some disruption of sleep. All participants visited the toilet at least once during the time they were meant to be asleep.  

\subsubsection{Work-related issues affect sleep}
Of the nine participants, three noted that work pressure may affect their sleep quality. These participants also noted that they worked more than eight hours a day with some working overtime and taking their work home. These people stated that they often found themselves lying on the bed, recalling the day's events and pondering their actions, concerned if they had made mistakes. Worrying about the next day's work was a similar problem, as P3 explains: \enquote{I'm just lying there, and that's when you then begin to reflect on what's going on during the day, what's going to be coming tomorrow...Obviously, your mind is just constantly being stimulated by these stressful thoughts.}

Some people admitted to checking their work emails on their smartphone just before going to bed which negatively affected their sleep: \enquote{you read that email before you go to bed, then you're up thinking about the email all night.} (P6). 

Some participants work shift cycles; this in itself can also lead to a lack of sleep due to an irregular routine. For example, employees who need to work at six in the morning will fall asleep but worry that they will oversleep or generally worry that they have little sleep to sleep during the daytime. If a person switches to a different shift within the same week, this can cause circadian clock disorder resulting in little sleep time between the last shift and first shift of a new cycle whilst the body readjusts. P7 stated: \enquote{[you] lying down in bed and settling down, maybe about 11:00pm and then you've got to be up at 5:00am, so that's like only six hours sleep still, between the two shifts.} Shift patterns make it increasingly difficult for a person to fall into a deep sleep: \enquote{so when you go to bed, it's not a deep sleep you get into. You're restless… but you're trying to force yourself to go to sleep.} (P8). Employees had previously indicated, through a work-based survey, that it is better to work in a fixed shift pattern: \enquote{People did a follow-up survey and they did say they were sleeping better by doing one shift rather than a rotating three-shift.}(P8).

Some of the work which has to get completed is time critical, such as the end of the financial year. This induces additional pressure (P1) along with other work-related obligations such as frequent travelling or \enquote{staying at the hotel} (P6), away from home in an unfamiliar sleeping environment. 

\subsubsection{Influence of workplace wellness program (Visibility of Data as Evidence)}
As already noted, the company is very conscious of employees mental and physical health. P5 states: \enquote{As a company, our objective would be for the welfare of the employees to make sure that they are not stressed...overstretched, and they're working an acceptable number of hours. And, when they leave work, they leave work at work, they don't take work home with them.} One of the nine participants believed that the sleep wellness could have a positive effect on their work performance: \enquote{It would improve sickness, reduce accidents, improve quality, improve output, and improve morale.}(P6).

The employer was viewed more favourably, even for allowing this study to take place as a first step towards developing a wellness program. Employees appreciated the thought of their company being receptive to their welfare: \enquote{I would say it's been good that they [the company] are doing this.} (P2).

However, four of nine participants expressed that they had underlying privacy concerns with the sleep tracking predominantly focused on sharing the data. For example, P2 stated: \enquote{sharing sleep data may cause others to question the life of people who sleep less, [...] or guess whether the employee's personal life will be troublesome or mentally ill.} 

All nine participants noted that using the sleep diary helped them record their behaviours and contributing factors that affect their sleep. People agreed that manually logging information was too time consuming and unsustainable in the long-term, expressing more interest in an automated IoT based solution that would be embedded within the workplace environment.  

\section{stage two design workshop and data visualisation}
IoT as independent federated service and a type of application architecture, which allows for automated data collection, acquisition, event transfer, network connectivity and interoperability \cite{Atzori2010Theinter}. Our IoT design workshop aimed to generate ideas around automating the tracking of sleep-related information (e.g. activities, behaviours and mental state). We then asked participants to imagine themselves working within the environment.  Our design outcomes reflect the participants perspective of what is suitable and acceptable tracking technology to be used within the workplace, as well as how employees would want to share their data. 

Many of our participants were not familiar with IoT devices, so we created small cards to help non-experts navigate the range of options that IoT devices offer. We adapted the existing 'Tiles IoT cards' toolkit \cite{Mora2017tiles} for this workshop. From this toolkit, we used the Mission (theme), Things, Human Action and Feedback cards as well as adapting our own Sensor and Data cards (please see figure 1). Things cards represent low-tech objects that you come into contact with on a daily bases, such as a coffee mug or a pencil. Human Action cards describe how the user can trigger a digital response from interacting with these objects, e.g. proximity sensor. Data cards comprise of factors that can affect sleep such caffeine intake, stress level and sleep duration. Mission cards outline a number of themes centred on human needs and desires and aim to trigger design thinking and creative dialogue amongst participants.

Additionally, two personas and four scenarios were used to provoke an alternative perspective on the problems highlighted in stage one's interview outcomes (e.g. automated ways of tracking hydration, exercise etc.).

We used two office floor plans printed on paper to simulate the working environment. The content of the workshop was as follows: 

\begin{itemize}
    \item Participants discuss the contents of the mission card and establish how to combine the sensor with the low-tech object (Thing) to collect data.
    \item Participants then marked the location of the IoT devices and dashboard on the floor plan. 
    \item Engage in discussion to decide which data should be kept personal, shared by name or remain anonymous.
    \item Depict graphics that visualise sleep and contributing factors. 
\end{itemize}

\subsection{Workshop findings}
In total, two identical workshops were run to accommodate participants work schedules. The first workshop saw 5 participants attend with 3 attending the second workshop. One person was unable to make either due to work-related commitments. Workshops were audio recorded, transcribed and then analysed using thematic analysis \cite{Baumer2015reflectiveinformatics}. All ideas generated from the IoT cards and floor plans were collected from the workshop; this material highlighted the extent to which technology could and should be used to automatically monitor employees sleep, exercise, food and water intake, feelings and stress levels. At this point, participants understood IoT technologies enough and were aware of the privacy issues associated with using them within a working environment. Employees were keen to point out that any use of IoT devices in the workplace should not be too intrusive and respect people's privacy. \enquote{The whole idea was to discuss what would be the most relevant, what would best suit us, obviously we got different jobs so all the applications of what they actually require will be different, so we thought about which one will be too invasive on your own personal life and which ones would be more ideal and how you could incorporate today's technology with the things that are happening in your life.} (P7). 

\subsubsection{Tracking and sharing the information related with sleep}
The user selected the appropriate sensors to complete the task on the mission card and marked out the locations where the sensors should be installed. Table 1 shows the outcomes.
\begin{table} 
\caption{The Location of Sensor Installation, Data Type and Sensor Type.}
 \label{tab:sensorloc}

\begin{tabular}{|p{2.2cm}|p{2.5cm}|p{2.5cm}|} 
\hline \textbf{Type of data} & \textbf{Location \& object} & \textbf{Equipment}\\
\hline Emotion, Stress and Feelings
  &Desk, or near Clock in machine
  &Camera, Fitness band(body movement, heart rate) \\
\hline Caffeine \& Water intake &Desk, Colour coded mug, bottles, toilet&
Activity sensor (accelerometer) strip, weight scale pad
Bluetooth scanner.
Bluetooth mug
Token-based coffee machine\\
\hline Diet type& Office&Mobile phone app(camera) \\
\hline Exercise&Human body&Fitness band, Smart phone\\
\hline Heart rate & body temperature &Human body, Fitness band \\
\bottomrule
\end{tabular} 
\end{table}
\subsubsection{Design for tracking daily subjective feelings }
Participants firstly discussed why they thought stress occurred and explained how they related that to their sleeping habits. They pointed out that they could use the webcam to track their daytime feelings, emotions and stress as well as food and water intake. P2 explained: \enquote{If you had your desktop camera it could recognise your face,[...]so it could see how you looked... It picks you up throughout the day…or you get to a point where you feel a little bit more lethargic, it could be your stress levels.}

Within our second workshop, one participant pointed out that food intake can be monitored by the company Closed-Circuit Television (CCTV) camera in the canteen. Later on, they change their mind as they thought cameras were too invasive as well as them being blocked by other objects or people. They then decided that tracking food intake through a mobile app which automatically recognised the food would be more suitable. 

\subsubsection{Privacy concerns of tracking in workplace}
From the workshops, our study found that users had different privacy considerations when using IoT devices for monitoring personal information. Employees expressed different levels of data granularity that they wanted to share; others did not want to share data at all, such as their emotional state as they believed it would lead to exposing past mental health issues: \enquote{sometimes it's easier in the office atmosphere, talking to somebody that you know but doesn't know your situation too well} (P2).

Participants highlighted that the use the cameras made people feel uncomfortable because \enquote{it's too invasive, too in your face type of thing and the microphone might be a little bit too invasive of your [...] personal conversations} (P2). Monitoring hydration was seen as a way of preventing participants from drinking too much water before going to bed. Some people wanted to record the volume of water drank during the day as well as the number of times a person went to the toilet. They thought this would be useful but other participants disagreed and pointed out that this also a too intrusive.

Participants were invited to design their wellness program data sharing system that would allow people to dictate which contributing factors they wanted to share with work colleagues and company management. Participants unanimously agreed that a total number of hours spent should be shared with the company by default.

\begin{figure*}
  \includegraphics[width=\textwidth,height=4cm]{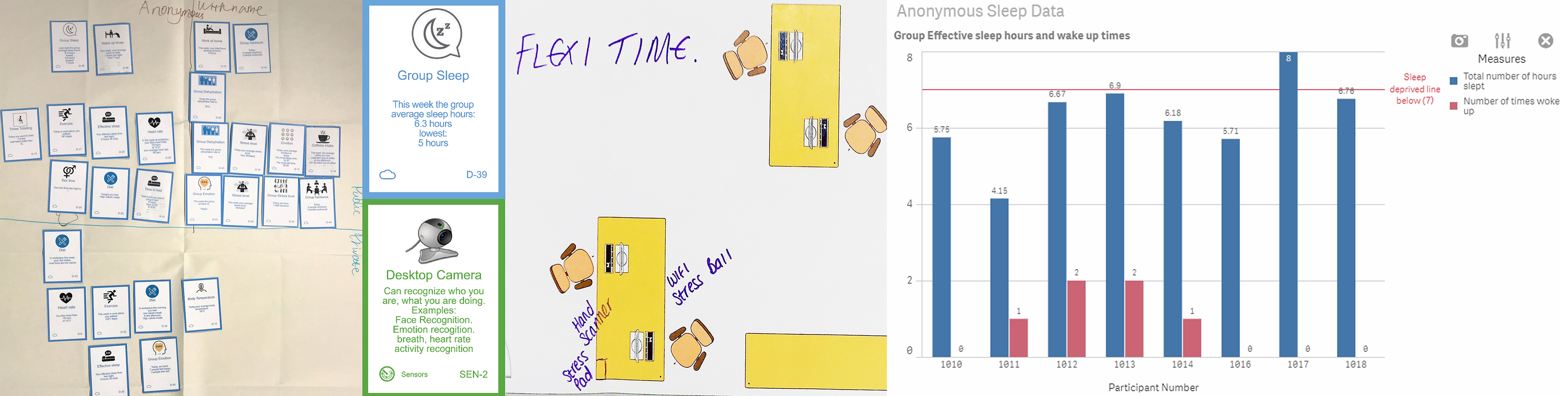}
  \caption{a)The outcome of the workshop, the top area is segmented by the data sharing level(anonymous, real name), the bottom section is personal data. b) two examples of IoT card, c) an example of an office floor plan, d) an example of a single nights anonymous sleep data dashboard}
\end{figure*}
\subsubsection{Unexpected use of data in workplace }
For the managers of the company, health and safety are at the forefront of their priorities. Managers believed that sleep not only impacted upon a person's physical and mental health but also directly related to workplace-based accidents. The responsibility and liability of an employee's safety fall on the company and the rules set out as part of occupational health guidelines. As P7 pointed out \enquote{if you come in tired you are more likely to do something daft and you are thinking, 'Why have I done that for?'}. P7 expressed a strong positive attitude towards tracking health indicators when examining Virgin's \cite{VirginPlus2018employee} wellness program. More importantly, P6 further pointed out a potential application of the data: \enquote{As a manager, if that data was being fed back to you then you know who is likely to have an accident on the shop floor ...How many of them are down to people being tired? ... Is it because he is tired? Is it complacency? Some of them will be tired, won't they?}

\subsection{Data sharing and depicting data visualisation }
In the final part of the workshop, attention turned to configuring how personal data should be shared within the workplace to make sleep deprivation more visible. Participants attempted to maintain a clear boundary between their private life and work. Within the discussion, the participants talked about the meaning of each Data card and drew boundaries on the paper. Cards were allocated into sections based on how the participant wanted the data to be shared. An example of how data was allocated can be seen in Figure 1.

\section{STAGE three DESIGN DASHBOARD PROBE}
The sleep data collected in stage one was analysed using Qlik\texttrademark; an online agile data visualisation tool that allows for the rapid building and deployment of interactive analytical apps (\url{www.qlik.com}). This allowed the participant to annotate the data by applying layers of filters to the data within the dashboard. 

The first set of dashboards displayed peoples personal sleep information such as total effective sleep time, total time lying on the bed, wake up times etc. The second set of dashboards highlighted the previous night's group sleep data and comprised of people's anonymous sleep data and their feelings that had been recorded within the sleep diary. Some participants' sleep-data was removed due to them changing their mind about sharing; the dashboard did still retained their subjective feelings. An example of the dashboards can be seen in Figure 1.

\subsection{Interview Findings: Interactive dashboards}
We conducted nine one-to-one audio recorded interviews (approximately 6.5 hours of recordings), selectively transcribed them and then analysed them thematically. Each participant was provide with a \pounds10 amazon voucher to compensate their time.

Each participant was asked to review the dashboards and then answer a set of questions designed to explore their understanding and any insights they had about the data. Open questions were used within the interviews to gain as much detail as possible, such as \enquote{Could you please pick up one or two of the charts and explain what you think they are telling you?}.

\subsubsection{Engaging with the data: discovering trends \& behavioural change}
Participants were presented with their sleep data through the dashboards. They noted that having a visualisation of the data helped to reinforce their accountability to themselves:\enquote{I think it's helpful to see things in black and white in front of you.}(P9) Sharing personal sleep data amongst colleagues allowed users to reflect on the group's data as P8 noted:\enquote{wow, I know people don't get enough sleep, but that is pretty bad.} Some participants felt responsible when comparing their own data with the group sleep data as P3 stated:\enquote{I think it's me pulling the group average down.}

Sharing data in an interactive and visual format can allow people to find similarities more accurately as well as help them learn by discovering patterns: \enquote{large scale you can find a more accurate pattern.}  (P7). For example, P4 found a relationship between their sleep and the type of food they had eaten by applying simple filters to their data: \enquote{A heavy meal means that they are sleeping less[hours] ...The light meal where you can clearly see that people are sleeping a lot better.} The dashboard allowed people to discover interesting themes and patterns associated with their own and group data: \enquote{different people will take different things out of it, but that's definitely useful}(P8).

Since the program began, some users started to take more notice of their daily water intake and began monitoring any changes. For example, \enquote{not having a coffee in the night now and trying to manage the level of the fluid I get} allowed P6 to reduce the number of times they woke up as well as the amount of times they visited the toilet.  

Overall, the one-to-one interviews highlighted that the majority of participants felt more accountable to themselves which is consistent with  The research finding is similar to \cite{Chung2017finding}.

\subsubsection{Fostering care for colleagues, and perceive care from the colleague and the employer}
The way in which data is shared can stimulate communication and a sense of care between people. For example, employees on shift work as a team: \enquote{you could be working with somebody who's getting no sleep [...] If you get this information [...] maybe you've got to have a bit more patience with him.} (P8). The visibility of data can also be used to draw attention to individual people who may be too shy or afraid to disclose a problem; other colleagues would sense the data irregularity, as P3 notes: \enquote{I think it's quite good if I made that public ...I am promoting people then to try and talk about my issue of lack of sleep.}

In launching the wellness program, the company not only benefits from the increased awareness of its employees physical and mental health, but it also reinforces the employee's perception that the company cares about them and their welfare. P8 believed that \enquote{things like this can only improve to me what companies consider to being their responsibility}. P1 think the wellness program can improve employee loyalty and further mentioned that \enquote{such projects usually happen in multinational companies}.

Other participants raised contrasting opinions over the need to share their sleep data, arguing it was strictly private: \enquote{sleep quality and sleep time is a very private topic.} (P3). Other participants expressed concerns over the awkwardness of asking for someone else's personal data, in that it may make them feel like they are being forced to expand or share their private space. \enquote{It's not work related...but close colleagues I think would probably question that} (P3). The willingness of participants to form new relationships is consistent with Chung et al. \cite{Chung2017finding} findings. 

\subsubsection{Sharing personal data within the workplace, and who owns it?}
People set boundaries that separate work from private life. Sleep is an activity that takes place outside of working hours. Amongst the participants from our workshops, 8 participants expressed no issues in sharing their sleep data. However, when participants were introduced to the dashboard they changed their opinion, especially when asked if they would share the data with their managers: \enquote{Only share if you have a good eight hours sleep... You've got nothing to hide, but if you did have a bad night's sleep, would you want your boss knowing?} (P9).

The remaining seven participants still maintained that sharing the data was acceptable and further believed that if the \enquote{company is going to take full responsibility for it, and want to help, then it's good, If they're just going to use it as a weapon against people, then this would obviously be an ethical issue. but I don't think I would be too bothered if the company knew} (P7).

Many people spend their weekends socialising and relaxing with their families. From our interviews, employees expressed that they do not want their employer recording their weekend sleep data, apart from on Sunday night, prior to starting work on Monday morning: \enquote{[...] which will mean they [company] are not thinking about your private life} (P3). In contrast, members of management believed that out-of-hours and weekend data should be included within the wellness program in an bid to educate employees about maintaining a consistent sleeping routine. 

Participants believed that data should be controlled by someone who is impartial. Some suggestions pointed to either the Information Technology (IT) team or Human Resources (HR) department. Management participants were keen to add that the governance of employee held data should be in the hands of the HR department, who should also be responsible for the management of the wellness program as an opt-in or opt-out program.

\subsubsection{Possible unexpected and problematical use of the wellness data}
When participants were asked if the company would use their data to treat their employees differently, P8 thought sleep should be outside the company's concern. It is \enquote{an individual's responsibility to turn up to work in a fit healthy state of mind}. However, this does not mean that the company has no responsibility at all; the company launched the wellness program and so they should consider themselves \enquote{as the whole umbrella...employee's welfare is a company responsibility} (P8).

Participants sleeping habits had a greater impact on work than other health-related activities. Managers primarily found sleep data more interesting as it could be used to prevent accidents. If an employee suffered acute sleep deprivation for one night (less than two or three hours sleep), the company would regard it as part of their sickness policy: \enquote{if it's a genuine reason then it would just be part of your sickness policy}(P6). In such a situation, the employee could be sent home to rest. Participants also highlighted that there more substantial sleep deprivation can occur as an effect of alcohol and drug intake. P6 further comments: \enquote{it is difficult to determine how many hours of sleep are enough} whilst participant explained that workers \enquote{had to be certain alertness to operate the machine ... it's health and safety ...it should be done}(P7).

P6 advocated that sleep data should be used as a way of exposing problems, rather than being used as a basis for disciplining employees: \enquote{seriously [you] have to look at your policies, but the first one will be education and the second one on there will be support. You wouldn't look to discipline your staff if somebody has a genuine issue}. P6 further explained that employees who have recently had children are likely subject to less sleep and should not be penalised for it.  In addition to the national maternity leave, the employer also considered adopting flexible working patterns to support employee's personal lives.

P7 highlighted that monitoring sleep data can also be used as a means of protecting employees: \enquote{go against the company... when There's no way really that they are going to be getting the average seven hours sleep}. 

Participants from management noted that people may potentially \enquote{forge the sleep figures} (P5) to make themselves appear as if they slept well; they proposed checking peoples sleep data on a quarterly basis, to monitor sleeping patterns over a longer time period. Another participant pointed out the system should prevent people from taking advantage of wellness data, such as \enquote{I want a day off... not have much sleep just so I can get sent home early}(P7) The findings here are similar to Gorm and Shklovski \cite{Gorm2016steps} who expressed concerns over morale and accuracy of data . 

One participant mentioned a potential solution to address the unexpected use of the sleep data when designing a wellness program: \enquote{you should consider consulting with the staff in policy making, and it is best for the management staff to train them separately. The best practice on [using their data], stay away of any workplace bullying [...] people need to be informed.} (P1) The participant also noted that a robust privacy policy should be in place: \enquote{make sure the data is not being negatively used}(P1). Furthermore, P1 thought there should be a policy \enquote{in place to have some sort of steps and have an opt-in, opt-out options as well [...] have the flexibility to back out, if the things aren't beneficial to the person or the company.}

\section{DISCUSSIONS AND DESIGN RECOMMENDATIONS}
Using digital technology as a tool to raise awareness of sleep deprivation is a laudable goal, and our study demonstrated that even a narrow focused, short-term research study can engage and enlighten people about the factors that influence their sleep, the importance of sleep quality and the length of sleep. Our findings highlight the complexities encountered when considering the design of a workplace-based sleep wellness program, the possible ramifications of raising awareness of sleep deprivation within the workplace as well as people's concerns over data ownership and governance. In light of these, we propose directions for the implementation of workplace based wellness programs that support mutual responsibility, trust and control of employee sleep data. 
\subsection{Co-designing a workplace-based wellness program}
Involving employees at the start of the design process was hugely beneficial in orientating people around the wellness project. It allowed for a gentle introduction into the concept of sleep deprivation, a concern that was not widely understood or acknowledged, as well as then clarifying what personal data is; how it can be collected using IoT devices and finally, establishing what they want to share on an individual basis, and the methods for sharing it within the workplace. By allowing people to take control of their personal data, privacy issues became a fairly minor concern that only surfaced when directly asked about them. This is in direct contrast with more recent concerns from both researcher and media \cite{Mattke2013workplace,Robinson2015thequantified,Van2003thecumulative}, where privacy concerns are at the forefront of people's discussion agenda. Only four of the nine participants expressed concerns around their data privacy when asked directly; this is consistent with other studies findings\cite{cooke2017sleep}.

Including employees in the process instilled a notion of transparency which helped to build trust; employees were initially sceptical towards the use of their data. Mutual trust was a factor that echoed throughout our findings at multiple points, it stands out as the core requirement in ensuring the sustainability of such a program. The employer's perspective of trust extended to wanting employees to engage with the program in a meaningful and fair way, with the aim of improving their lifestyle based on the program; they also wanted employees to track as much of their sleep as possible, even on weekends. Employees regarded trust as non-bias, fair and impartial use of personal data by their employer so that it could not be used as weapon against them at any point. 

Evidently, work-based wellness programs need to take careful consideration of how they are implemented. Consideration needs to be given to the employees and the context of the program. For example, our manufacturing industry-based study may not be applicable to that of a software development company; enforcing a blanket approach across different environments will not ensure employee uptake. Furthermore, we recommend that employers plan ahead, past the implementation stage,  to consider how they can further support employees, as opposed to simply monitoring them. Ultimately, this is a space which requires further exploration beyond our study. 

\subsection{Raising awareness of sleep deprivation}
As much published literature has noted, wellness programs may cause privacy concerns when sharing personal data in workplace\cite{cooke2017sleep,Green2007peer,Wells2017incorporating}. Press and media frequently cast a negative perception on the ever-evolving nature of workplace-based monitoring. Despite negative coverage, there have been a number of deployments which have explored sleep and fatigue monitoring as well as people's subjective feelings \cite{Doherty2017largescale,oconnor2015wearables,Rooksby-2014-personaltracking}.

There were strong concerns over the potential mistreatment of people's personal sleep data. People are uncomfortable at the thought of their employer analysing their sleeping routines as it could lead to employees being perceived differently or even being bullied because of their routines. Introducing personal data into a more open environment has the potential to reveal people's previously hidden health issues, for example, a disability, depression or insomnia; if these details were exposed, this could potentially lead to dismissal from the workplace. Alternatively, if an employer examined someone's personal data, it is unclear where the responsibility and liability fall if a problem were to be identified from that person's sleeping routine. Would the employer intervene as a duty of care or would the responsibility shift to the employee to monitor their own personal data. 

Introducing the sleep wellness program into the workplace has encouraged dialogue between employees as well as prompting an increase in employee-employer discussion around personal data. By using the interactive dashboards, people could see others data; it would be possible to engage with others and express concern or care over their well-being; this would be especially beneficial for individuals who are more introverted. Moreover, the dashboards provide a learning platform for employees to examine useful information by querying their data. The dashboards created an interactive environment that allows employees to monitor their sleep in a peer-supported and self-driven manner, similar to Digital Civics' relational services\cite{Olivier2015digital}. From an employer's perspective, the dashboards allowed for the monitoring and checking of employee statuses,however the data also provides evidence and context should an employee call in sick; data can be used to provide a snapshot timeline and some evidential insight into a person's most recent routine to help inform the authorisation of sick-days.

Through exploring the visibility of sleep data within the workplace, there are positive and negatives outcomes; encouraging a supportive work environment will allow people to more openly share their information. However, this requires a systemic change within the workplace which must be supported by robust policies that provide mutual understanding and trust, for both employees and employers alike. 

\subsection{Data ownership and governance}
We previously highlighted how co-design provided us with a means of establishing the way in which employees wanted to share their data. Employees and employers alike were unsure as to who should have ownership of people's private data, if it was gathered using company bought equipment, as well as being collected during working hours. Employers were keen to maintain access to workers data for the purposes of fatigue risk management. This is due to effects of sleep deprivation being similar to in-taking alcohol\cite{Roehrs1994sleepiness,Wells2017incorporating}, which would lead to significant health and safety concerns within the workplace. The amount of time data is held as well as who has access to it were prominent concerns of employees who outlined that guidelines need to be in place to control access. Emerging block-chain technology offers a means of negating data placement concerns, as data can be held within network nodes as part of a larger decentralised system. Smart contracts\cite{nissen2018geocoin,christidis2016blockchains} would offer token-based access to private data. Both Employees and employers would be segregated from the data but governance could be mediated through one-time access tokens whereby rules are established for how each party can access, use and analyse the data.  

Data can be interpreted in a multitude of ways; private data could be analysed at a different point in time, using different tools and algorithms. This has the potential of revealing much more than an employee initially expected. Acquiring employees consent for how private data will be analysed is critical to building trust within the program. Not only should consent be obtained in the first instance, but private data should be considered longitudinal whereby consent is re-applied for from the employer\cite{cockburn2018hark} should alternative ways of interpreting it surface \cite{bowyer2018understanding}. Such explicit identification of how data will be used have existed for a long time within the field of medicine. 

We would expect there to be defined policies within the workplace that explicitly state how data is used, who owns it, who can access it as well as who controls it and how long it is retained; similar to how the General Data Protection Regulation (GDPR) functions. Adapting the consent-based practices of the medical field would also be a good starting point.

\section{CONCLUSION \& FUTURE WORK}
This study presents a series of design considerations that should be integrated into a workplace-based wellness program when co-designing alongside employees. We highlight employees concerns surrounding the use of their personal data, how it is shared as well as highlight their fears of how it can be misused by employers; we propose a series of guidelines for employers to implement sleep deprivation awareness programs in order to ensure mutual responsibility and trust.  

Future work should seek to clarify the ownership of data within the workplace environment as well as outline the distribution of governance employers have over their employee's personal data. Further case studies should apply more formalised and established policies surrounding workplace-based wellness programs so that longer term deployments can highlight additional factors contributing to program sustainability, employee retention, effectiveness, usability and acceptability. These further studies will help to mould future platforms that can empower employees to take ownership of their own data and control how it is shared amongst other employees within the workplace.

%
\begin{acks}
This research was funded through the EPSRC Centre for Doctoral
Training in Digital Civics (EP/L016176/1). Data supporting this
publication is openly available under an 'Open Data Commons
Open Database License'. Please visit {\url{https://github.com/famousgrouse/pervasivehealth}} for access instructions.
\end{acks}

%
\bibliographystyle{ACM-Reference-Format}
\bibliography{sleep}

%

\end{document}